\documentclass[aps,floatfix,nofootinbib]{revtex4}
\usepackage{graphicx}

\begin{document}

\title{Non conducting spherically symmetric fluids}
\author{Kayll Lake \cite{email}}
\affiliation{Department of Physics, Queen's University, Kingston,
Ontario, Canada, K7L 3N6 }
\date{\today}

\begin{abstract}
A class of spherically symmetric spacetimes invariantly defined by
a zero flux condition is examined first from a purely geometrical
point of view and then physically by way of Einstein's equations
for a general fluid decomposition of the energy-momentum tensor.
The approach, which allows a formal inversion of Einstein's
equations, explains, for example, why spherically symmetric
perfect fluids with spatially homogeneous energy density must be
shearfree.
\end{abstract}
\maketitle

Any spacetime is an ``exact solution" Einstein's equations, but
only those which are ``physically reasonable" are worthy of
consideration. Whereas the term ``exact solution" has no precise
definition, the concept is well understood \cite{kshm}
\cite{krasinski}. The concept of ``physically reasonable" is
somewhat more slippery and it is the responsibility of the author
of any new exact solution to justify its significance. Clearly the
force fitting of a metric to a sufficiently general
energy-momentum tensor is but nonsense. However, proceeding from a
general energy-momentum tensor to more restricted cases can
provide insight into properties of the idealizations. Here we
consider spherically symmetric spacetimes which admit a fluid
source subject to a zero flux condition equivalent to the
requirement of zero thermal conductivity. First, we proceed in an
invariant purely geometrical fashion. Then, by way of Einstein's
equations, we translate our results into physical (albeit
phenomenological) parameters.
\bigskip

Consider a spherically symmetric spacetime $\mathcal{M}$
\cite{notation}
\begin{equation}
ds^2_{\mathcal{M}}=ds^2_{\Sigma}+R^2d\Omega^2
\end{equation}
where $d\Omega^2$ is the metric of a unit sphere
($d\theta^2+sin^2(\theta)d\phi^2$) and $R=R(x^1,x^2)$ where the
coordinates on the Lorentzian two space $\Sigma$ are labelled as
$x^1$ and $x^2$. (No specific choice of coordinates on $\Sigma$
will be made.) Consider a flow (a congruence of unit timelike
vectors $u^{\alpha}$) tangent to an open region of $\Sigma$.
Writing $n^{\alpha}$ as the normal to $u^{\alpha}$ in the tangent
space of $\Sigma$ we restrict $\mathcal{M}$ by the condition that
there exists $u^{\alpha}$ such that
\begin{equation}
G_{\alpha}^{\beta}u^{\alpha}n_{\beta}=0, \label{flux}
\end{equation}
where $G_{\alpha}^{\beta}$ is the Einstein tensor of
$\mathcal{M}$. (The Vaidya metric \cite{vaidya} (corresponding to
a null flux) provides a familiar example of an $\mathcal{M}$ for
which such a flow does not exist.) Now $u^{\alpha}$ (and
$n^{\alpha}$) are, if they exist, uniquely determined from the
metric alone. (Equations $u^{\alpha}u_{\alpha}=-1$,
$n^{\alpha}n_{\alpha}=1$, $u^{\alpha}n_{\alpha}=0$ and the zero
flux condition (\ref{flux}) are solved for the components of
$u^{\alpha}$ and $n^{\alpha}$ and the appropriate roots determined
by the time orientation of $\mathcal{M}$ (say $u^1>0$).)  The
algorithm for producing $u^{\alpha}$ for specific coordinates on
$\Sigma$ is straightforward and discussed elsewhere in a more
general context \cite{mus}. In what follows we make use of the
following scalars:
\begin{equation}
G \equiv  G_{\alpha}^{\alpha}, \; G1 \equiv
G_{\alpha}^{\beta}u^{\alpha}u_{\beta}, \; G2 \equiv
G_{\alpha}^{\beta}n^{\alpha}n_{\beta}, \;
\Delta\equiv\sigma_{\alpha}^{\beta}n^{\alpha}n_{\beta},\;\Psi
\equiv G+G1-3G2,\;\mathcal{W}\equiv C_{\alpha \beta \gamma
\delta}C^{\alpha \beta \gamma \delta}, \label{scalars}
\end{equation}
where $\sigma_{\alpha}^{\beta}$ is the shear associated with
$u^{\alpha}$ and $C_{\alpha \beta \gamma \delta}$ is the Weyl
tensor. A necessary and sufficient condition for the conformal
flatness of $\mathcal{M}$ is $\mathcal{W}=0$.
\bigskip

Perhaps the most important function associated with $\mathcal{M}$
is the effective gravitational mass $M(x^1,x^2)$ the invariant
properties of which were first explored by Hernandez and Misner
\cite{hm} who wrote the function in the form
\begin{equation}
M= \frac{R^{3}}{2}\mathcal{R}_{\theta \phi}^{\; \; \;\; \theta
\phi} , \label{mass}
\end{equation}
where $\mathcal{R}$ is the Riemann tensor of $\mathcal{M}$. See
also \cite{cm}, \cite{pe}, \cite{z} and \cite{h} for further
discussion and \cite{visser} for a recent application. In terms of
the scalars (\ref{scalars}) $M$ can be given as
\begin{equation}
M= \frac{R^3}{12}(G+3(G1-G2)+\sqrt{3\mathcal{W}}) \label{massnew}
\end{equation}
from which we immediately obtain the following necessary and
sufficient condition for the conformal flatness of $\mathcal{M}$:
\begin{equation}
M= \frac{R^3}{12}(G+3(G1-G2)). \label{flatnew}
\end{equation}
From (\ref{flux}) and (\ref{massnew}) it follows that
\begin{equation}
M^{'}= \frac{G1}{2}R^2R^{'} \label{mprime}
\end{equation}
where $^{'}$ is the intrinsic derivative in the direction
$n^{\alpha}$. (For the case of a perfect fluid in comoving
coordinates (\ref{mprime}) is well known (e.g. \cite{krasinski}
section 1.4).) Assuming that $\mathcal{M}$ is non-singular (the
scalars (\ref{scalars}) remain finite) and has an origin ($R=0$)
it follows from (\ref{massnew}) and (\ref{mprime}) that
\begin{equation}
G1^{'}=0 \Leftrightarrow G+G1-3G2+\sqrt{3\mathcal{W}}=0
\label{G1prime}
\end{equation}
or, equivalently,
\begin{equation}
G1^{'}=0 \Leftrightarrow M=\frac{G1 R^3}{6}. \label{G1primemass}
\end{equation}
Observe that if
\begin{equation}
\Psi=0 \label{psiflat}
\end{equation}
then
\begin{equation}
G1^{'}=0 \Leftrightarrow \mathcal{W}=0 \label{psiflat1}.
\end{equation}

\bigskip
To transform the forgoing invariant geometrical statements into
physical phenomenological parameters by way of Einstein's
equations, decompose the energy-momentum tensor in the form
\begin{equation}
T^{\alpha}_{\beta}=\rho
u^{\alpha}u_{\beta}+p_{1}n^{\alpha}n_{\beta}
+p_{2}\delta^{\alpha}_{\beta} +
p_{2}(u^{\alpha}u_{\beta}-n^{\alpha}n_{\beta})-2\eta
\sigma^{\alpha}_{\beta}. \label{imperfect}
\end{equation}
Einstein's equations are effectively reversed by way of the
following observation: Substitution of
\begin{equation}
\rho=\frac{G1}{8 \pi}, \label{rho}
\end{equation}
\begin{equation}
p_1=\frac{G2}{8 \pi}+2 \eta \Delta, \label{p1}
\end{equation}
and
\begin{equation}
p_2=\frac{G+G1-G2}{16 \pi}-\eta \Delta, \label{p2}
\end{equation}
into (\ref{imperfect}) and multiplication by $8 \pi$ produces
$G_{\alpha}^{\beta}$. Note in particular that there are at most
three independent Ricci scalars in this case and that the syzygy
(17) of \cite{pollney} for the decomposition (\ref{imperfect})
must, by way of the foregoing inversion, reduce to an identity as
can be directly verified \cite{mus}. The following specializations
arise, where in all cases the energy density is given by
(\ref{rho}):
\bigskip

\paragraph{$\eta \Delta=0$: }
If $\eta \Delta=0$ (one or more of $\eta, \Delta$ or
$\sigma_{\alpha}^{\beta}=0$) then $\Psi=0$ is a necessary and
sufficient condition for a perfect fluid ($p_1=p_2, \eta=0$).
\bigskip

\paragraph{$p_1 \equiv p_2 \equiv p,\; \Delta \neq 0$: }
If $p_1 \equiv p_2 \equiv p,\; \Delta \neq 0$ it follows from
(\ref{p1}) and ({\ref{p2}) that
\begin{equation}
p=\frac{G+G1}{24 \pi},
\end{equation}
and
\begin{equation}
\eta=\frac{G+G1-3G2}{48 \pi \Delta},
\end{equation}
so that $\Psi=0$ is once again a necessary and sufficient
condition for a perfect fluid.
\bigskip

\paragraph{$p_1 \neq p_2,\; \Delta\neq 0$: }
Now $p_1$ is given by (\ref{p1}) and $p_2$ by (\ref{p2}) where
$\eta$ is a freely specified function. The special choice $\eta
\equiv 0$ reduces to case a.
\bigskip

Clearly any spacetime $\mathcal{M}$ satisfying (\ref{flux}) is an
``exact solution" of Einstein's equations: simply put (\ref{rho}),
(\ref{p1}) and (\ref{p2}) into (\ref{imperfect}).  Only with the
imposition of a constraint, like $\Psi=0$, is $\mathcal{M}$
restricted. For example, in the static case, where
$\sigma_{\alpha}^{\beta}=0$, any $\mathcal{M}$ is a ``solution",
but with the constraint $\Psi=0$ relatively few available
solutions make any physical sense \cite{dl} \cite{new}. As
emphasized previously, only those ``exact solutions" which make
physical sense are worthy of consideration. The possibility that a
given $\mathcal{M}$ might have more than one physically reasonable
interpretation is not excluded, but all possible interpretations
subject to (\ref{imperfect}) are delineated above. Because the
foregoing procedure is algorithmic, and because alternate
interpretations of know solutions are of some interest if
physically justified, the procedure is being incorporated into the
interactive database GRDB \cite{computer}.

\bigskip
In terms of the parameters $(\rho,p_1,p_2,\eta)$ we have
\begin{equation}
M= \frac{R^3}{12}(16 \pi
(\rho+p_2-p_1+3\eta\Delta)+\sqrt{3\mathcal{W}}),
\label{massnewparm}
\end{equation}
\begin{equation}
\mathcal{W}=0 \Leftrightarrow M=\frac{4
\pi}{3}R^3(\rho+p_2-p_1+3\eta\Delta), \label{flatnewparm}
\end{equation}
\begin{equation}
G1^{'}=0 \Leftrightarrow M=\frac{4 \pi}{3}R^3\rho,
\label{G1primemassrho}
\end{equation}
and if
\begin{equation}
p_2-p_1+3\eta\Delta=0 \label{descriminant}
\end{equation}
then
\begin{equation}
G1^{'}=0 \Leftrightarrow \mathcal{W}=0.\label{descriminant1}
\end{equation}
Condition (\ref{descriminant1}) generalizes a previously known
condition \cite{lake}. A perfect fluid is a special case of
(\ref{descriminant}) and in that case we have
\begin{equation}
G1^{'}=0 \Leftrightarrow \mathcal{W}=0 \Rightarrow
\sigma^{\alpha}_{\beta}=0 \label{descriminant2}
\end{equation}
where the last implication comes from the fact that all
conformally flat perfect fluids are shear free \cite{steph}. The
condition $G1^{'}=0 \Rightarrow \sigma^{\alpha}_{\beta}=0$, for a
perfect fluid in the comoving frame, is stated in \cite{kshm}
(section (14.2.4)) and attributed to \cite{tw} and \cite{ms}.
Neither \cite{kshm} (in that section), \cite{tw} nor \cite{ms}
even mention the Weyl tensor, the vanishing of which, as the
foregoing makes clear, can be considered central to the result.
\begin{acknowledgments}
This work was supported by a grant from the Natural Sciences and
Engineering Research Council of Canada. Portions of this work were
made possible by use of \textit{GRTensorII} \cite{grt}. It is a
pleasure to thank Don Page and Nicos Pelavas for comments. A more
complete account of related work, but with a different emphasis,
is in progress with Mustapha Ishak.
\end{acknowledgments}

\end{document}